

\font\titolino=cmbx10
\font\tsnorm=cmr10
\font\tscors=cmti10

\font\tscorsp=cmti9
\magnification=1200

\hsize=148truemm
\hoffset=10truemm
\parskip 3truemm plus 1truemm minus 1truemm
\parindent 8truemm
\newcount\notenumber

\def\2d{two-\-di\-men\-sio\-nal }
\def\3d{three-\-di\-men\-sio\-nal }
\def\4d{four-\-di\-men\-sio\-nal }

\def\PRD{{\tscors Phys. Rev. D }}
\def\NPB{{\tscors Nucl. Phys. B }}

\def\PLB{{\tscors Phys. Lett. B }}

\def\PRP{{\tscors Phys. Rep. }}
\def\JMP{{\tscors J. Math. Phys. }}

\def\MPLA{{\tscors Mod. Phys. Lett. A  }}
\def\CQG{{\tscors Class. Quantum Grav. }}

\def\CMP{{\tscors Commun. Math. Phys. }}
\def\GRG{{\tscors Gen. Rel. Grav. }}
\def\mpl{M_{pl}^2}
\def\d{\partial}

\def\sc{\scriptstyle}

\def\note{\advance\notenumber by 1 \footnote{$^{\the\notenumber}$}}
\def\ref#1{\medskip\everypar={\hangindent 2\parindent}#1}
\def\beginref{\begingroup
\bigskip
\leftline{\titolino References.}
\nobreak\noindent}
\def\endref{\par\endgroup}
\def\ra{\rightarrow}
\def\beginsection #1. #2.
{\bigskip
\leftline{\titolino #1. #2.}
\nobreak\noindent}
\def\beginappendix
{\bigskip
\leftline{\titolino Appendix.}
\nobreak\noindent}
\def\beginack
{\bigskip
\leftline{\titolino Acknowledgments.}
\nobreak\noindent}

\nopagenumbers
\null
\vskip 5truemm
\rightline {INFNCA-TH-94-11}
\rightline {SISSA 75/94/A}
\rightline{June 1994}
\vskip 15truemm
\centerline{\titolino COSMOLOGICAL AND WORMHOLE SOLUTIONS}
\bigskip
\centerline{\titolino IN LOW-ENERGY EFFECTIVE STRING THEORY}
\vskip 15truemm
\centerline{\tsnorm Mariano Cadoni$^{(a),(c)}$ and Marco
Cavagli\`a$^{(b),(d)}$}
\bigskip
\centerline{$^{(a)}$\tscorsp Dipartimento di Scienze Fisiche,}
\smallskip
\centerline{\tscorsp Universit\`a  di Cagliari, Italy}
\bigskip
\centerline{$^{(b)}$\tscorsp Sissa - International School for Advanced
Studies,}
\smallskip
\centerline{\tscorsp Via Beirut 2-4, I-34013 Trieste, Italy.}
\bigskip
\centerline{$^{(c)}$\tscorsp INFN, Sezione di Cagliari,}
\smallskip
\centerline{\tscorsp Via Ada Negri 18, I-09127 Cagliari, Italy.}

\bigskip
\centerline{$^{(d)}$\tscorsp INFN, Sezione di Torino, Italy.}
\vskip 15truemm
\centerline{\tsnorm ABSTRACT}
\begingroup\tsnorm\noindent
We derive and study a class of cosmological and wormhole solutions of
low-energy effective string field theory. We consider a general
four-dimensional string effective action where moduli of the compactified
manifold and the electromagnetic field are present. The cosmological
solutions of the two-dimensional effective theory obtained by dimensional
reduction of the former are discussed. In particular we demonstrate that
the two-dimensional theory possesses a scale-factor duality invariance.
Euclidean four-dimensional instantons describing the nucleation of the
baby universes are found and the probability amplitude for the nucleation
process is given.
\vfill
\leftline{\tsnorm PACS: 11.25.-w, 11.25.Hj, 04.20.Jb\hfill}
\smallskip
\hrule
\noindent
\leftline{E-Mail: CADONI@CA.INFN.IT\hfill}
\leftline{E-Mail: CAVAGLIA@TSMI19.SISSA.IT\hfill}
\endgroup
\vfill
\eject
\footline{\hfill\folio\hfill}
\pageno=1
\beginsection 1. Introduction.
During the last years low-energy  effective string field theory has been
widely investigated in connection with its black hole [1-6] and
cosmological [7-9] solutions. In particular critical and non-critical
string cosmological solutions have been discovered and analysed in
detail. The typical features of these solutions such as the scale-factor
duality or the relationship with the corresponding black hole geometries
have been also investigated [7-9]. The general idea underlying these
investigations is that the short-distance modifications of string theory
to general relativity could be crucial in order to understand
long-standing problems of quantum gravity such as the loss of information
in the black hole evaporation process or the nature of the singularities
in the Einstein theory. After all, string theory is the only consistent
framework which we presently have for quantizing gravity and it seems
very natural to look at it in order to solve these problems.

If the gravitational field is correctly described by a quantum theory -
as string theory -, the topology of spacetime is expected to fluctuate on
Planck length scales. This is the old idea of {\tscors spacetime foam}
[10]. Microscopic connections between large regions of spacetime
(wormholes, in the following WHs) and the chaotic formation of Planck
scale universes (baby universes, in the following BUs) branching off, or
joining onto, a region of spacetime can have important effects on
physics, even at low energy scales [11]. Semiclassically, WHs are
described by instantons which represent a tunneling between two asymptotic
\4d regions of spacetime. If a WH can be joined at the throat to a
hyperbolic universe whose spatial section is compact, then the instanton
can be interpreted as nucleating a BU with a well defined semiclassical
amplitude.

Even though a large number of Euclidean instantons - and the corresponding
cosmological solutions describing BUs - have been found in the context
of the  Einstein gravity theory [12-14], up to now very little is known
about
WHs and BUs nucleation in the framework of the low-energy effective string
theory. A first attempt in this direction was made by Giddings et al. in
ref. [12]. Using a lowest order effective string action they found an
axionic WH which however has an infinite Euclidean action.
More recently an instanton has been found which describes a tunneling
between zero-energy vacua of string theory [15].
This situation
is somehow uncomfortable because if string theory has to solve the
puzzles of quantum gravity it should also provide us the natural
framework for studying processes like the formation of WHs and BUs.

In this paper we shall derive and study \4d cosmological solutions
of the low-energy  string effective  action of ref. [5,6]. This action
takes into account, apart from the dilaton and the EM field,
a modulus field which acquires non minimal couplings to the gauge fields
owing to string one-loop effects. Furthermore it contains, as particular
cases, both the dilaton-gravity action of ref. [2,3] and the
Einstein-Maxwell action. These cosmological solutions describe
universes filled with the EM and the dilaton field. Even
though these solutions do not seem to correspond, at least in the general
case, to exact conformal string backgrounds, they are interesting from
two different points of view. First, the dimensional reduction of the
\4d theory on the background defined by the magnetically charged
solutions produces a \2d effective theory whose solutions have all the
features of \2d string cosmological solutions. Second, the \4d
cosmological solutions can be interpreted as BUs nucleated by
Euclidean WHs.

The outline of the paper is the following. In the next section we find
and discuss the cosmological solutions of the \4d action with both a
purely magnetic and electric field. In sec. 3 we study the \2d effective
theory obtained by dimensional reduction of the \4d one. In particular
we demonstrate that this theory possesses a scale-factor duality
invariance and that its time-dependent solutions describe the region
between the horizon and the singularity of the corresponding black hole
geometry. In sect. 4 we find the Euclidean instantons of the \4d theory.
These solutions can be joined at $t=0$ to their analytic continuations in
the hyperbolic spacetime described by the cosmological solutions of sec.
2 and then we are led to interpret the instantons as nucleating BUs. The
probability amplitude for  this process is also calculated. Finally we
state our conclusions in sec. 5.
\beginsection 2. Cosmological solutions.
Let us consider a \4d hyperbolic manifold $\Omega$ with
metric $g_{\mu\nu}$ and topology $R\times H$, where $H$ is a
\3d compact Riemannian hypersurface with metric $h_{ij}$
(here and in the following greek indices run from $0$ to $4$ and latin
indices run from $1$ to $3$). Our starting point is the \4d
low-energy string effective action of ref. [5,6] which generalizes the usual
low-energy string effective action [2,3,16] for the case when a modulus is
taken into account. Owing to string one-loop threshold effects this
modulus is coupled to the EM field.

The action reads:
$$\eqalign{S=\int_\Omega d^4x\sqrt{|g|}e^{-2\phi}\biggl[R+4(\nabla\phi)^2&
-{2\over 3}(\nabla\psi)^2- F^2-
e^{2\phi-(2/3)q\psi}F^2\biggr]+\cr\cr
&+2\int_{\d\Omega}
d^3x \sqrt{h}e^{-2\phi}({\bf K}-{\bf K_0}).\cr}\eqno(2.1)$$
Here $R$ is the curvature scalar, $\phi$ is the dilaton field, $\psi$ a
modulus field, $F_{\mu\nu}$ is the usual EM field
tensor and $q$ is a coupling constant. We have
put $16\pi G\equiv\mpl/16\pi=1$, then measuring all dimensional quantities
in these units. The boundary term is required by unitarity [17]; ${\bf K}$ is
the trace of the second fundamental form ${\bf K}_{ij}$ of
$H$ and $\bf{K_0}$ is that of the asymptotic
three-surface embedded in flat space. The latter contribution must
be introduced if one requires the spacetime to be asymptotically flat.
We will see that the surface term in (2.1) plays a very important role in
the computation of the probability amplitude for BU formation.

Following [5,6] we choose for the modulus field the ansatz
$$e^{-(2/3)q\psi}={3\over q^2}e^{-2\phi}.\eqno(2.2)$$
Using (2.2), the action (2.1) reduces to the form
$$\eqalign{S=\int_\Omega d^4x\sqrt{|g|}e^{-2\phi}
\biggl[R&-{\sc{8k}\over\sc{1-k}}
(\nabla\phi)^2-{\sc{3+k}\over\sc{1-k}}F^2\biggr]\cr\cr
&+2\int_{\d\Omega}
d^3x \sqrt{h}e^{-2\phi}({\bf K}-{\bf K_0}),}\eqno(2.3)$$
where
$$k={3-2q^2\over 3+2q^2},\qquad -1\le k\le 1.\eqno(2.4)$$
We have several interesting cases according to the value of $k$. For
$k=-1$ (i.e. $q\ra\infty$) the action reduces to the usual low-energy
string action when the modulus $\psi$ is not taken into account; (2.3)
describes then the \4d dilaton-gravity theory considered in
ref. [2,3]. For $k=0$ we have a \4d action whose \2d
reduction gives the Jackiw-Teitelboim theory [6]. The case $k=1$ looks
singular. However, inserting $q=0$ in the action (2.1) and using the
equations of motion that force the dilaton to be constant, we recover the
usual Einstein-Maxwell theory. Using the ansatz (2.2) exact solutions can
be obtained for any value of the coupling constant $k\in [-1,1]$.

Since the EM field prevents spatially homogenous and isotropic
solutions of the field equations, we look for solutions of the form
$$ds^2=-N^2(t)dt^2+a^2(t)d\chi^2+b^2(t)d\Omega_2^2,\eqno(2.5)$$
where $\chi$ is the coordinate of the one-sphere, $0\le\chi<2\pi$, and
$d\Omega_2^2$ represents the line element of the two-sphere. $N(t)$ is
the lapse function. The line element (2.5) is known as of the
Kantowski-Sachs type [18] and describes a hyperbolic spacetime $\Omega$
whose \3d spatial hypersurfaces $H$ have topology $S^1\times
S^2$.

Now, we have to consider a form of the EM field compatible with the
topology of the spacetime. A suitable configuration is given by the
magnetic monopole on the two-sphere:
$$F=Q_m\sin{\theta}d\theta\wedge d\varphi,\eqno(2.6)$$
where $Q_m$ is the magnetic charge. Eq. (2.6) describes a purely magnetic
field. Later on this section we will consider  a purely electric field
with  only non-vanishing component along the $\chi$ direction [14]:
$$A=A(t)d\chi,\eqno(2.7)$$
where $A$ is the EM potential one-form. As we shall see, the solutions
corresponding to (2.6) and (2.7) are related by a duality transformation.

It is straightforward (but not so easy...see Appendix) to find the
magnetic charged solution of the equations of motion derived from the
action (2.3):
$$\eqalignno{&ds^2=-dt^2+Q^2\sin^2{(t/Q)}
\biggl[{1+\cos{(t/Q)}\over 2}\biggr]^{k-1}
d\chi^2+Q^2d\Omega_2^2,&\hbox{(2.8a)}\cr\cr
&e^{2(\phi-\phi_0)}=
\biggl[{1+\cos{(t/Q)}\over 2}\biggr]^{(k-1)/2},&\hbox{(2.8b)}\cr\cr}$$
where we have redefined the magnetic charge $Q_m$ through
$$Q_m={1\over 2}\sqrt{1-k} \, Q.\eqno(2.9)$$
The previous solution exists and is well defined for any $-1\le k<1$. For
$k=1$ the redefinition of the magnetic charge (2.9) becomes singular.
This is not surprising because the ansatz (2.2) is singular for $k=1$
(i.e. $q=0$); so, the solutions for this particular case have to be
determined by starting directly from the action (2.1) with $q=0$. One can
easily verify that the corresponding solution is described by (2.6-8)
where now $k=1$ and $Q_m=Q$.

Let us study the properties of the solution (2.8). The line element
(2.8a) describes a universe whose spatial sections are compact with
topology $S^1\times S^2$. The scale factor of the two-sphere is constant,
while the radius of the one-sphere is periodic in time. The behaviour of
the line element (2.8a) depends on $k$,  so it is convenient to study
separately the following cases:

\item{{\tscors a})} $k=1$, the Einstein-Maxwell theory. In this case the
line element reduces to the one found in [14] for the Einstein
gravity. The radius $a$ of the one-sphere takes values in the range
$[0,Q]$ and the line element is singular at $t=n\pi Q$, $n=0,\pm 1,\pm
2...$, where the scale factor $a$ vanishes. This singularity can be
removed by a different choice of coordinates [14]. This can be easily seen
noting that in the neighbourhood of $t=0$ the line element (2.8a) reduces
to the form
$$ds^2=-dt^2+t^2d\chi^2+Q^2d\Omega_2^2.\eqno(2.10)$$
The topology is locally $R^{1,1}\times S^2$ and the \3d
spatial hypersurface becomes homotopic to $S^2$ and a point. Thus (2.8a)
represents a universe which periodically reproduces itself with period
$\pi Q$. In this case the dilaton is constant.

\item{b)} $0<k<1$. Contrary to the previous case, when $k$ takes values
in the interval $]0,1[$, the metric has a curvature singularity for
$t=(2n+1)\pi Q$ where the dilaton diverges and the theory becomes
strong-coupled. At $t=2n\pi Q$ there is a coordinate singularity
analogous to the case a). The radius of the one-sphere vanishes for
$t=n\pi Q$ and has a maximum for $\cos(t/Q)=(k-1)/(k+1)$. In this case
(2.8a) describes a universe whose two-sphere scale factor remains
constant, whereas the radius of the one-sphere vanishes at $t=0$, grows
till a maximum value and becomes again zero after a time $t=\pi Q$.

\item{c)} $k=0$. In this case (2.8a) describes a periodic universe with
period $2\pi Q$. The scale factor $a$ vanishes for $t=2n\pi Q$, where
there is a coordinate singularity analogous to the case a), and takes its
maximum value $a_{max}=2Q$ when $t=(2n+1)\pi Q$. Note that even though
there are no curvature singularities, the dilaton diverges and the theory
becomes strong-coupled for $t=(2n+1)\pi Q$.

\item{d)} $-1\le k<0$. The scale factor $a$ vanishes for $t=2n\pi Q$,
where the metric shows a coordinate singularity, and goes to the infinity
for $t=(2n+1)\pi Q$ where the line element has a curvature singularity.
Hence the radius of the one-sphere starts with zero at $t=0$ and grows to
infinity at $t=\pi Q$. Note that $k=-1$ corresponds to the usual
dilaton-gravity theory.

As we shall see in sec. 4, $\forall k\in [-1,1]$ the solution (2.8) can
be joined at $t=0$ with an Euclidean asymptotically flat instanton and
then (2.8a) can be interpreted as the line element of a BU nucleated
starting from an asymptotically flat region.

To conclude this section, let us briefly discuss  the solutions obtained
from the purely electric field (2.7). It is well known that the equations
of motion for  \4d dilaton gravity coupled to the EM
field are invariant under the discrete duality transformation [3]
$$F_{\mu\nu}\to {1\over 2}e^{-2\phi}\epsilon_{\mu\nu}{}{}^{\lambda\rho}
F_{\lambda\rho}, \qquad \phi\to-\phi,\qquad g_c\to g_c,\eqno(2.11)$$
when expressed in terms of the canonical metric $g_c=e^{-2\phi}g_s$. The
transformation (2.11) relates magnetically to electrically charged
solutions. This invariance also holds for the theory described by (2.3)
when the action is expressed in terms of the canonical metric $g_c$ [6].
Using the duality invariance (2.11) of the equations of motion, it is
straightforward to obtain in the string frame the following electrically
charged solution:
$$\eqalignno{&ds^2=e^{4\phi_0}\biggl[{1+\cos{(t/Q)}\over 2}\biggr]^{1-k}
\biggl\{-dt^2+Q^2\sin^2{(t/Q)}\cdot&\cr
&\hbox to 3truecm{}\cdot\biggl[{1+\cos{(t/Q)}\over 2}\biggr]^{k-1}
d\chi^2+Q^2d\Omega_2^2\biggr\},&\hbox{(2.12a)}\cr\cr
&F={1\over 2}\sqrt{1-k}~
e^{2\phi_0}\sin{(t/Q)}dt\wedge d\chi,&\hbox{(2.12b)}\cr\cr
&e^{2(\phi-\phi_0)}=
\biggl[{1+\cos{(t/Q)}\over 2}\biggr]^{(1-k)/2},&\hbox{(2.12c)}\cr\cr}$$
where $Q$ is related to the electric charge by the same relation as in
(2.9). Note that for $k=1$ (2.8a) and (2.12a) coincide after taking
$\phi_0=0$; indeed, in this
case the dilaton is constant and the duality invariance holds also in the
string frame. The solution (2.12) has properties analogous to the
solution (2.8). The line elements (2.8a) and (2.12a) differ only for a
conformal factor because of the duality relation. The most striking
difference between the two solutions resides in the fact that differently
from (2.8) the scale factor for the two-sphere in (2.12a) is not
constant. As we shall see in the following, the \2d section
of the magnetic solution (2.8) can be described in terms of an effective
\2d theory obtained by dimensional reduction of the action
(2.3). This is of course not possible for the electric solution (2.12).

\beginsection 3. The \2d effective theory and dual solutions.
The metric part of the magnetic solution (2.8) has the form of a direct
product of a \2d solution and a two-sphere of constant
radius. Thus, it is useful to study the \2d effective
theory obtained by retaining only the time-dependent modes of the
\4d theory. This \2d theory is expected to
describe the essential \4d physics for perturbations around
the background solution (2.8). The action (2.3) can be dimensionally
reduced by taking the angular coordinates to span a two-sphere of
constant radius  $Q$. The resulting \2d action is
$$S=\int d^2x\sqrt{|g|}e^{-2\phi}
\biggl[R-{\sc{8k}\over\sc{1-k}}
(\nabla\phi)^2+\lambda^2\biggr],\eqno(3.1)$$
where $\lambda^2=(1-k)/2Q^2$. This \2d action has been
studied in connection with its black hole solutions and its duality
invariances in ref. [6,19]. In this section we will study (3.1) from the
cosmological point of view. As shown in ref. [19] considering
space-dependent field configurations, the action (3.1) possesses a
duality symmetry. It is easy to see that this duality invariance also
holds for time-dependent configurations. Let us consider the metric and
the dilaton field of the form
$$ds^2=-dt^2+e^{2\rho(t)}dx^2, \qquad \phi=\phi(t),\eqno(3.2)$$
where $0\le x<\infty$. The action becomes
$$S=\int dt e^{-2\phi+\rho}\bigl[2(\ddot\rho+\dot\rho^2)+
{\sc{8k}\over \sc{1-k}}\dot\phi^2+\lambda^2\bigr],\eqno(3.3)$$
where the dots represent time-derivatives.
One can easily check that the transformation:
$$\rho\to k\rho -2(k+1)\phi,\qquad \phi \to {k-1\over 2}\rho
-k\phi\eqno(3.4)$$
leaves the action invariant modulo a total derivative. The duality
transformation (3.4) is the generalization for the action (3.1) of the
scale-factor duality symmetry of string theory [7,9]. Indeed, for
$ k=-1$ we get the standard scale-factor duality transformation
$ \rho\to -\rho,\quad \phi\to
\phi -\rho$ which exchanges the radius of the \2d universe
with its inverse.

Let us now discuss the cosmological solutions of the \2d
theory and their behaviour under the duality transformation (3.4). The
time-dependent solution of the action (3.1) is easily found to be:
$$\eqalignno{&ds^2=-dt^2+ \sin^2{(t/2Q)}\bigl[\cos^2{(t/2Q)}\bigr]^{k}
dx^2,&\hbox{(3.5a)}\cr\cr
&e^{2(\phi-\phi_0)}=\bigl[\cos^2{(t/2Q)}\bigr]^{(k-1)/2}
.&\hbox{(3.5b)}\cr\cr}$$
Considering a periodic space, i.e setting $x=2Q\chi, 0\le
\chi<2\pi,$ the solution (3.5) coincides with the \2d section of
solution (2.8). The effect of the duality transformation (3.4) on the
solution (3.5) is to exchange the sine and cosine everywhere:
$$\eqalignno{&ds^2=-dt^2+ \cos^2{(t/2Q)}\bigl[\sin^2{(t/2Q)}\bigr]^{k}
dx^2,
&\hbox{(3.6a)}\cr\cr
&e^{2(\phi-\phi_0)}=\bigl[\sin^2{(t/2Q)}\bigr]^{(k-1)/2}.
&\hbox{(3.6b)}\cr\cr}$$
The dual solution corresponds of course to a solution of the
\4d theory (see Appendix). Moreover, for $k=1$ (3.5) and
(3.6) are the same, i.e the solution is self-dual. Comparing eq. (3.5)
with eq. (3.6) and keeping in mind the discussion of the previous
section, one easily realizes that the effect of the duality
transformation (3.4) on the solutions with $k\not=1,0$ is to exchange the
coordinate singularities at $t=2n\pi Q$ with the curvature singularities at
$t=(2n+1)\pi Q$. For $k=0$ there are no curvature singularities and the
duality transformation simply exchanges strong string couplings with weak
ones.

The previous cosmological solutions are a further example of the
\2d string cosmologies studied in ref. [7-9]. They exhibit all the
peculiar properties of string cosmological solutions such as the
above-discussed duality invariance.
In particular for $k=-1$ the solutions (3.5) and (3.6) correspond
to well-known $D=2$ cosmological conformal string backgrounds [7-9].
However for generic $k$ we do not know if the interpretation of
(3.5) and (3.6) as conformal string backgrounds can be maintained.
 In this context the case $k=0$ seems
very interesting. As we have seen, the cosmological solution describes an
universe which periodically reproduces itself without encountering a
singularity, thus avoiding the singularity-problem which affects the
models with $k\not= 0,1$.

To conclude this section, let us discuss the relationship between the
\2d cosmological solutions and the corresponding
\2d black hole geometries. For the particular case $k=-1$, it
has been already shown that the cosmological solution (3.5) describes the
region between the horizon and the singularity of the black hole
geometry [9] derived from (3.1)
$$\eqalignno{&ds^2=-4Q^2 \tanh^2{(x/2Q)}d\tau^2
+dx^2,
&\hbox{(3.7a)}\cr\cr
&e^{2(\phi-\phi_0)}=\bigl[\cosh{(x/2Q)}\bigr]^{-2}.
&\hbox{(3.7b)}\cr}$$
This construction can be easily generalized for arbitrary $k$. Consider
the metric (3.5a) expressed in terms of the periodic coordinate
$\chi=x/2Q$ and choose the new coordinates:
$$\eqalign{u=e^{t^* +\chi},&\qquad v=e^{t^*-\chi},\cr\cr
t^*={1\over2Q}\int{dt\over\sin{(t/2Q)}\cos^k{(t/2Q)}}&=
{y^{(1-k)/2}\over k-1}
{\it F}\biggl({1-k\over2},1,{3-k\over 2},y\biggr),\cr\cr}\eqno(3.8)$$
where ${\it F}$ is the hypergeometric function and $y=\cos^2{(t/2Q)}.$
The line element (3.5a) becomes
$$ds^2= -4Q^2{(1-y)y^k\over uv}du dv.\eqno(3.9)$$
An identical form for the metric can be obtained starting from the black
hole solution of the action (3.1) [6]:
$$\eqalignno{&ds^2=-4Q^2
\sinh^2{(x/2Q)}\bigl[\cosh{(x/2Q)}\bigr]^{2k}d\tau^2
+dx^2,
&\hbox{(3.10a)}\cr\cr
&e^{2(\phi-\phi_0)}=\bigl[\cosh{(x/2Q)}\bigr]^{k-1}
,&\hbox{(3.10b)}\cr}$$
and introducing the coordinates:
$$\eqalign{u=e^{x^* +\tau},&\qquad v=e^{x^*-\tau},\cr\cr
x^*={1\over2Q}\int{dx\over\sinh{(x/2Q)}\cosh^k{(x/2Q)}}&=
{y^{(1-k)/2}\over k-1}
{\it F}\biggl({1-k\over2},1,{3-k\over 2},y\biggr),\cr}\eqno(3.11)$$
but now with $y=\cosh^2(x/2Q)$. In (3.8) $y$ takes values in the interval
$[0,1]$, whereas in (3.11) $1\le y<\infty$. Hence the solution (3.9)
describes the region between the horizon and the singularity of the black
hole solution (3.10). Strictly speaking, because we are working with
$\chi$ periodic, this correspondence holds for a wedge in the region
between the horizon and the singularity (see [9]).

As shown in ref. [9] for the case $k=-1$,  one could continue the
time past the singularity
at $t=\pi Q$ to get an identical copy of the interior of the
black hole where the universe now starts at the singularity
evolves till it reaches zero size at $t=2\pi Q$.
By continuing this procedure, i.e. not identifying
$t\to t+2\pi Q$ we end up with an universe which undergoes
infinitely many oscillations.
This construction  cannot be taken too seriously because near the
singularity, where the size of the universe becomes infinitely large,
one cannot trust anymore the low-energy string effective action (3.1) and
one should consider the exact theory.
We will not discuss this point further, we just note that our model
with $k=0$ avoids the singularity problem. Indeed, for this value
of $k$ the scalar curvature stays
everywhere finite and only the dilaton diverges at $t=(2n+1)\pi Q$
indicating that the theory becomes  strong-coupled.

\beginsection 4. Euclidean instantons.
Let us now discuss the action (2.3) in the Riemannian space. In order to
do this, we have to deal with the ambiguity in the choice of the sign for
the action of the EM field. Indeed, the EM field in the Euclidean space
is not analytically related to the EM field in the hyperbolic space,
i.e. the two fields are not related by the coordinate transformation
$t\ra i\tau$: a real electric field in the hyperbolic spacetime gives,
once continued in the Euclidean space, an imaginary electric field. Using
a more or less ad hoc rule, one normally requires the Euclidean EM field
and its action $S_E$ to be real, i.e one uses the hyperbolic expression
of the action with an overall minus sign (the discussion about the
Euclidean formulation of the Maxwell theory can be found in ref. [20]).
Of course, other prescriptions are possible. We will consider the
Euclidean version of the action (2.3) in the following general form:
$$\eqalign{S_E=\int_\Omega d^4x\sqrt{|g|}e^{-2\phi}
\biggl[-R&+{\sc{8k}\over\sc{1-k}}
(\nabla\phi)^2+\varepsilon{\sc{3+k}\over\sc{1-k}}F^2\biggr]\cr\cr
&-2\int_{\d\Omega}
d^3x \sqrt{h}e^{-2\phi}({\bf K}-{\bf K_0}),}\eqno(4.1)$$
where $\varepsilon=\pm 1$. The meaning of the parameter $\varepsilon$ can
be understood looking at the EM field. According to the sign of
$\varepsilon$, the electric and the magnetic fields in the hyperbolic and
Euclidean space are related by
\medskip\noindent
\line{\hfill$E^2_{\rm hyp}=\varepsilon E^2_{\rm Eucl}$,\hfill
$H^2_{\rm hyp}=-\varepsilon H^2_{\rm Eucl}$.\hfill}
\medskip\noindent
Thus, if we require the analytical continuations of real hyperbolic fields
to be real fields in Euclidean space, we must choose $\varepsilon=-1$ for
the purely magnetic configuration (2.6) and $\varepsilon=1$ for the
purely electric configuration (2.7). Another argument that supports the
above choice relies on the duality invariance (2.11). One can easily see
that (2.11) does not hold for the Euclidean theory. This is essentially
due to the fact that the Euclidean EM energy is proportional to
$\exp(-2\phi) (E^2-H^2)$, thus changing sign under the transformation
(2.11). Hence, in order to maintain the duality relation (2.11) in the
Euclidean space we must reverse the sign of $\varepsilon$ in the action
in passing from the purely magnetic configuration (2.6) to the purely
electric configuration (2.7).

Let us first consider the purely magnetic ansatz. We have to choose
$\varepsilon=-1$ in (4.1) and the Euclidean equations of motion are solved by:
$$\eqalignno{&ds^2=d\tau^2+2^{(1-k)}Q^2{\tau^2\over\tau^2+Q^2}
\biggl(1+{Q\over\sqrt{\tau^2+Q^2}}
\biggr)^{k-1}d\chi^2+&\cr\cr
&\hbox to 5truecm{}+(\tau^2+Q^2)d\Omega_2^2,&\hbox{(4.2a)}\cr\cr
&F={\sqrt{1-k}\over 2}Q\sin{\theta}d\theta\wedge
d\varphi,&\hbox{(4.2b)}\cr\cr
&e^{2(\phi-\phi_0)}=2^{(1-k)/2}
\biggl(1+{Q\over\sqrt{\tau^2+Q^2}}\biggr)^{(k-1)/2}.&\hbox{(4.2c)}\cr}$$
Let us study the properties of the solution (4.2). In the asymptotic
regions $\tau\ra\pm\infty$,  the line element becomes
$$ds^2=d\tau^2+2^{(1-k)}Q^2d\chi^2+\tau^2d\Omega_2^2.\eqno(4.3)$$
Thus the asymptotic Riemannian space is flat with topology $R^3\times
S^1$. At $\tau=0$, $\forall k \in [-1,1]$, the metric is singular. This
singularity is only due to the choice of the coordinates that cover
only half of the manifold described by the line element (4.2a).
Indeed, it is possible to find a new chart that covers the whole space
(for the case $k=1$, see [14]). One can easily verify this,
observing that in the neighbourhood of $\tau=0$, (4.2a) becomes
$$ds^2=d\tau^2+\tau^2d\chi^2+Q^2d\Omega_2^2,\eqno(4.4)$$
hence the singularity at $\tau=0$ can be removed going to Cartesian
coordinates in the $(\tau,\>\chi)$ plane and adding the point
$\tau=0$. Thus, in the neighbourhood of $\tau=0$ the topology is
locally $R^2\times S^2$ with $R^2$ contracting to zero as $\tau\ra 0$.
This particular case has been classified by Gibbons and Hawking [21] as
a ``bolt'' singularity.

The asymptotic behaviour of solution (4.2) and its regularity allow us to
interpret the instanton (4.2a) as a WH that connects two asymptotic flat
regions. Moreover, the instanton (4.2) can be joined at $\tau=0$ with the
cosmological solution (2.8). Thus (4.2) describes  the nucleation of a
BU starting from an original flat region. Let us see this in detail.

As one can easily see, solutions (4.2) with $-\infty<\tau<0$ and (2.8)
with $t>0$ satisfy Darmois conditions for change of signature at
$t=\tau=0$ [22]; indeed, at $t=\tau=0$ both the Euclidean and hyperbolic
manifolds are well defined and the first and second fundamental forms of
the \3d hypersurfaces coincide smoothly for $\tau\ra 0^-$
and $t\ra 0^+$. The dilaton is continuous with its derivative on the
hypersurface $t=\tau=0$, where the change of signature occurs.
Therefore, the asymptotic behaviour for $\tau\ra\pm\infty$ of the
Euclidean solution (4.2) allow us to interpret (4.2) as an instanton
which provides a tunneling between a flat vacuum region and the universe
described by (2.8a).

In conclusion, solution (4.2) describes the nucleation of a non isotropic
BU at $t=0$ starting from an original flat spacetime and it is the
generalization to effective string theory of the solution found in [14]. The
hypersurface of signature change is \2d: this
corresponds to the particular situation of a BU nucleated in a phase of
maximum shrinkage of the spatial metric. Once the BU is nucleated then
it evolves according to (2.8) and eventually ends in a singularity
depending on the parameter $k$.

Let us now discuss the amplitude probability of nucleation of a BU. In
the semiclassical approximation, the amplitude probability in a Planck
volume and in a Planck time is given by [23]:
$$\Gamma=e^{-|\tilde S|},\eqno(4.5)$$
where $\tilde S$ represents the Euclidean action evaluated on the
solutions of the classical equations of motion.
After a straightforward calculation and taking into account
the boundary terms to cancel the divergent contribution coming
from the asymptotic region, one finds
$$\Gamma=\exp{[-8\pi^2 e^{-2\phi_0}Q^2(k+1)]}.\eqno(4.6)$$
For $k\not=-1$, in order to have a probability of the order of unity, the
charge $Q$ appearing in the solution must be of the order of the unity,
so the nucleation probability is maximum for BUs with dimension of order
of the Planck length. Conversely, for the usual dilaton-gravity theory
($k=-1$), the semiclassical amplitude probability (4.6)  does not fix the
dimension of the BU, because one obtain $\Gamma=1$ for any value of the
charge $Q$. In this case, in order to fix the amplitude probability one
must consider higher order contributions in the string tension
$\alpha'$ to the low-energy string effective action.

To conclude the section, let us discuss the Euclidean instanton for the
purely electric field (2.6). Now we have to choose $\varepsilon=-1$
in the action (4.1) and  we find:
$$\eqalignno{&ds^2=e^{4\phi_0}\biggl(1+{Q\over\sqrt{\tau^2+Q^2}}
\biggr)^{1-k}\biggl[d\tau^2+&\cr
&\hbox to 1truecm{}2^{(1-k)}Q^2{\tau^2\over\tau^2+Q^2}
\biggl(1+{Q\over\sqrt{\tau^2+Q^2}}
\biggr)^{k-1}d\chi^2+
(\tau^2+Q^2)d\Omega_2^2\biggr],&\hbox{(4.7a)}\cr\cr
&F={1\over 2}\sqrt{1-k}
Q^2e^{2\phi_0}{\tau\over\bigl(\tau^2+Q^2\bigr)^{3/2}}d\tau\wedge
d\chi,&\hbox{(4.7b)}\cr\cr
&e^{2(\phi-\phi_0)}=2^{(k-1)/2}
\biggl(1+{Q\over\sqrt{\tau^2+Q^2}}\biggr)^{(1-k)/2}.&\hbox{(4.7c)}\cr}$$
Analogously to the purely magnetic configuration, one can easily verify that
the solution (4.7) can be joined at $t=\tau=0$ to the electric
cosmological solution (2.12), so (4.7) can be interpreted as an instanton
nucleating a BU with metric (2.12a). The amplitude probability for this
process coincides with (4.6).
\beginsection 5. Conclusions.
In this paper we have derived and studied a class of cosmological
solutions for low-energy effective string field theory. These \4d
solutions describe universes filled with the dilaton and a purely
magnetic or a purely electric field. They are characterized by a
parameter $k$ and their behaviour depend crucially on this parameter. In
particular, in the purely magnetic case, for $k=1$ (the Einstein-Maxwell
theory) and for $k=0$, the solution describes a universe which
periodically reproduces itself without running in a singularity. For
$k\not=0,1$ the solution describes a universe which ends in a
singularity. This behaviour holds in particular for $k=-1$ which is the
case of the usual dilaton gravity coupled to the EM field. We have also
shown as the theory can be dimensionally reduced to a \2d theory. The \2d
theory exhibits a scale-factor duality symmetry which is a generalization
of the duality symmetries found for exact cosmological conformal string
backgrounds.  For $k=-1$ the \2d solution reduces to a well known string
conformal background [7-9]. Finally, the \2d  time-dependent solutions
describe the region between the horizon and the singularity of the black
hole solutions of the \2d theory. This feature has been found also for exact
conformal string backgrounds.

The \4d cosmological solutions can be also interpreted as BUs nucleated
starting from a flat spacetime region. The Euclidean instantons describing
this process have been found and the amplitude probability of nucleation
has been calculated.
\beginack
We wish to thank S. Mignemi for interesting remarks.
\beginappendix
Here we deduce the solution (2.8) and its dual of sec. 3. Since
computations are not trivial, we shall show them in detail. The
calculations for the Euclidean solutions of sec. 4 are analogous.

Our starting point is the action (2.3). The calculation is much simpler
if we express the action in terms of the canonical metric. Rescaling the
metric as $g_{\mu\nu}~\ra~\hbox{exp}(2\phi)~g_{\mu\nu}$ we get the action
in the canonical frame:
$$\eqalign{S=\int_\Omega d^4x\sqrt{|g|}
\biggl[R&-2{3+k\over 1-k}\biggl(
(\nabla\phi)^2+{1\over 2}e^{-2\phi}F^2\biggr)\biggr]\cr\cr
&+2\int_{\d\Omega}
d^3x \sqrt{h}({\bf K}-{\bf K_0}).}\eqno\hbox{(A.1)}$$
The equations of motion are:
$$\eqalignno{&G_{\mu\nu}=2{3+k\over 1-k}\biggl[\nabla_\mu\phi
\nabla_\nu\phi-{1\over
2}g_{\mu\nu}(\nabla\phi)^2+e^{-2\phi}\biggl(F_{\mu\rho}F_{\nu}{}^\rho-{
1\over 4}g_{\mu\nu}F^2\biggr)\biggr],&(\hbox{A.2a})\cr\cr
&\nabla^2\phi=-{1\over 2}e^{-2\phi}F^2,&(\hbox{A.2b})\cr\cr
&\nabla_\mu\bigl(e^{-2\phi}F^{\mu\nu}\bigr)=0.&(\hbox{A.2c})\cr}$$
Using the ansatz (2.6) for the EM field, we can easily see that the eq.
(A.2c) is identically satisfied. Substituting the line element (2.4) and
(2.6) in the other equations, (A.2) reduce to
$$\eqalignno{&{\dot b^2\over b^2}+{N^2\over b^2}+2{\dot a\dot b\over
ab}={3+k \over 1-k}\biggl(\dot\phi^2+Q_m^2{N^2\over b^4}
e^{-2\phi}\biggl),&\hbox{(A.3a)}\cr\cr
&2{\ddot b\over b}+{\dot b^2\over b^2}-2{\dot b\dot N\over bN}+
{N^2\over b^2}=-{3+k \over 1-k}\biggl(\dot\phi^2-Q_m^2{N^2\over b^4}
e^{-2\phi}\biggl),&\hbox{(A.3b)}\cr\cr
&{\ddot b\over b}+{\ddot a\over a}+{\dot a\dot b\over ab}-{\dot bN\over
bN}-{\dot aN\over aN}=
-{3+k \over 1-k}\biggl(\dot\phi^2+Q_m^2{N^2\over b^4}
e^{-2\phi}\biggl),&\hbox{(A.3c)}\cr\cr
&{d~\over dt}\biggl({ab^2\over N}\dot\phi\biggr)=Q_m^2{Na\over
b^2}e^{-2\phi},&\hbox{(A.3d)}\cr}$$
where the dot represents the derivative with respect to the time. The
Lagrangian density in the minisuperspace is (we neglect the surface terms
since these  do not affect the equations of motion):
$$L=2\biggl[-{a\dot b^2\over N}-{2\dot a b\dot b\over N}+Na+
{3+k\over 1-k}\biggl({ab^2\over N}\dot\phi^2-{Na\over b^2}
Q_m^2e^{-2\phi}\biggl)\biggr].\eqno\hbox{(A.4)}$$
Solving the equations of motion in the form (A.3) is very hard. Eqs.
(A.3) and the Lagrangian density can be greatly simplified using the new
variables:
\par\noindent
\centerline{\hfill $N=e^{2\rho+\nu},$\hfill $a=a_0e^\nu,$\hfill
$b=b_0e^\rho$.\hfill(A.5)}
\par\noindent
The variables defined in (A.5) reduce the system to a Toda-lattice
form [3]. Thus, defining $\xi=\nu+\rho$ and neglecting an overall constant
factor, the Lagrangian density (A.4) becomes:
$$L=-\dot\xi^2+\dot\nu^2+{e^{2\xi}\over b_0^2}+{3+k\over 1-k}
\biggl(\dot\phi^2-{Q_m^2\over b_0^4} e^{2(\nu-\phi)}\biggl).
\eqno\hbox{(A.6)}$$
Varying (A.6) with respect to $\xi$, $\nu$ and $\phi$ we obtain
the equations of motion for the new variables
$$\eqalignno{&\ddot\xi=-{e^{2\xi}\over b_0^2},&\hbox{(A.7a)}\cr\cr
&\ddot\nu=-{3+k\over 1-k}
{Q_m^2\over b_0^4}e^{2(\nu-\phi)},&(\hbox{A.7b})\cr\cr
&\ddot\phi={Q_m^2\over b_0^4} e^{2(\nu-\phi)}.&\hbox{(A.7c)}\cr}$$
The Hamiltonian constraint is
$$\dot\xi^2-\dot\nu^2+{e^{2\xi}\over b_0^2}=
{3+k\over 1-k}\biggl(\dot\phi^2+{Q_m^2\over b_0^4}
e^{2(\nu-\phi)}\biggl)=0.\eqno\hbox{(A.8)}$$
Now, we are able to solve (A.7-8). From (A.7a) we obtain
$$\xi=\log{(\alpha b_0)}-\log\hbox{cosh}~\bigl[\alpha(
t-t_0)\bigr]\eqno\hbox{(A.9)}$$
where $\alpha$ and $t_0$ are  integration constants. From
(A.7b) and (A.7c), defining $\chi=\nu-\phi$, we have
$$\eqalignno{&\phi={1-k\over 4}(\gamma t-\chi)+\delta,&\hbox{(A.10a)}\cr
&\nu={1\over 4}\bigl[(1-k)\gamma
t+(3+k)\chi\bigr]+\delta,&\hbox{(A.10b)}\cr
&\chi=\log\beta-\log\hbox{cosh}~\bigl[\beta(
t-t'_0)\bigr]-{1\over 2}\log\biggl[{4Q_m^2\over(1-k)b_0^4}\biggr],
&(\hbox{A.10c})\cr}$$
where $\beta$, $\gamma$, $\delta$, and $t'_0$ are integration
constants.
Substituting (A.9) and (A.10) in the Hamiltonian constraint, we find that
$\alpha$, $\beta$ and $\gamma$ must satisfy the relation
$$\alpha^2-{1-k\over 4}\gamma^2-{3+k\over 4}\beta^2=0.\eqno(\hbox{A.11})$$
Choosing $\alpha=\beta=\pm\gamma$ (A.11) is satisfied $\forall k\in
[-1,1]$. Recalling (A.5) and setting $t_0=t'_0=0$ we find
$$\eqalignno{&N=\alpha Q  e^{-\phi_0}e^{\mp
\alpha t(1-k)/4}\bigl[\hbox{cosh}~(\alpha
t)\bigr]^{-(5-k)/4},&\hbox{(A.12a)}\cr\cr
&a=a_0'e^{-\phi_0}e^{\pm
\alpha t(1-k)/4}\bigl[\hbox{cosh}~(\alpha
t)\bigr]^{-(3+k)/4},&\hbox{(A.12b)}\cr\cr
&b=Q e^{-\phi_0}e^{\mp
\alpha  t(1-k)/4}\bigl[\hbox{cosh}~(\alpha
t)\bigr]^{-(1-k)/4},&\hbox{(A.12c)}\cr\cr
&e^{\phi-\phi_0}=e^{\pm
\alpha t(1-k)/4}\bigl[\hbox{cosh}~(\alpha
t)\bigr]^{(1-k)/4},&\hbox{(A.12d)}\cr}$$
where $a_0'$ and $\phi_0$ are arbitrary constants and $Q$ is given in
terms of $Q_m$ by eq. (2.9). Choosing the positive sign and rescaling the
metric to the string frame, (A.12a-c) become
$$\eqalignno{&N=\alpha Q \bigl[\hbox{cosh}~(\alpha
t)\bigr]^{-1},&\hbox{(A.13a)}\cr\cr
&a=a_0'e^{
\alpha t(1-k)/2}\bigl[\hbox{cosh}~(\alpha
t)\bigr]^{-(1+k)/2},&\hbox{(A.13b)}\cr\cr
&b= Q.&\hbox{(A.13c)}\cr}$$
 To express the line element as a  function of the proper time we
perform the coordinate transformation
$$\tau=2 Q~\hbox{arctg}~\bigl(e^{\alpha
t}\bigr).\eqno\hbox{(A.14)}$$
The solution becomes
$$\eqalignno{&ds^2=-d\tau^2+a_0'^2\sin^2{(\tau/Q)}\bigl[1+\cos {(\tau/Q)
}\bigr]^{k-1}d\chi^2+
Q^2 d\Omega^2_2,&\hbox{(A.15a)}\cr\cr
&e^{2(\phi-\phi_0)}=\bigl[1+\cos{(\tau/Q)
}\bigr]^{(k-1)/2}.&\hbox{(A.15b)}\cr}$$
{}From (A.15) we get  (2.8) with a suitable choice of the constant $a_0'$.
Finally, we would emphasize that choosing $\gamma=-\alpha$ we obtain the
\4d dual solution corresponding to (3.6).
\vfill\eject
\beginref
\ref [1] E. Witten, \PRD {\bf 44}, 314 (1991).

\ref [2] D. Garfinkle, G.T. Horowitz and A. Strominger, \PRD {\bf 43},
3140 (1991);

\ref [3] G.W. Gibbons and K. Maeda, \NPB {\bf 298}, 741 (1988).

\ref [4] A. Giveon, \MPLA {\bf 6}, 2843 (1991).

\ref [5] M. Cadoni and S. Mignemi, \PRD {\bf 48}, 5536 (1993).

\ref [6] M. Cadoni and S. Mignemi, \NPB (in press), hep-th 9312171.

\ref [7] G. Veneziano, \PLB {\bf 265}, 287 (1991);
         M. Gasperini and G. Veneziano \PLB {\bf 277}, 265 (1992).

\ref [8] M. Mueller, \NPB {\bf 337}, 37 (1990).

\ref [9] A. A. Tseytlin and C. Vafa, \NPB {\bf 372}, 443 (1992).

\ref [10] S.W. Hawking, \NPB {\bf 144}, 349 (1978).

\ref [11] S.W. Hawking, \NPB {\bf 335}, 155 (1990) and references therein.

\ref [12] S.B. Giddings and A. Strominger, \NPB {\bf 306}, 890 (1988).

\ref [13] R.C. Myers, \PRD {\bf 38}, 1327 (1988); J.J. Halliwell and R.
La\-flam\-me, \CQG {\bf 6}, 1839 (1989); S.W. Hawking, \PLB {\bf 195}, 337
(1987); A. Hosoya and W. Ogura, \PLB {\bf 225}, 117 (1989); B.J. Keay and
R. Laflamme, \PRD {\bf 40}, 2118 (1989).

\ref [14] M. Cavagli\`a, V. de Alfaro and F. de Felice, \PRD {\bf 49},
6493 (1994).

\ref [15] D. Brill and G. Horowitz, \PLB {\bf 262}, 437 (1991).

\ref [16] E. Witten,  \PLB {\bf 155}, 151 (1985).

\ref [17] S.W. Hawking, {\tscors in} General Relativity, an Einstein
Centenary Survey, eds. S.W. Hawking and W. Israel (Cambridge University
Press, Cambridge, 1979).

\ref [18] R. Kantowski, R. and R.K. Sachs, \JMP {\bf 7}, 443 (1966).

\ref [19] M. Cadoni and S. Mignemi, Report No: INFNCA-TH-94-4,
hep-th\-/9403113.

\ref [20] D. Brill, Report No: UMD 93-038, gr-qc/9209009, to appear
{\tscors in} Proceedings of Louis Witten Festschrift, World Scientific.

\ref [21] G.W. Gibbons and S.W. Hawking, \CMP {\bf 66}, 291 (1979); T.
Eguchi, P.B. Gilkey and A.J. Hanson, \PRP {\bf 66}, 213 (1980).

\ref [22] W.B. Bonnor and P.A.Vickers, \GRG {\bf 13}, 29 (1981); G.F.R.
Ellis and K. Piotrkowska {\tscors in} Proceedings of the
Journ\'ees Relativistes, Brussels, 1993.

\ref [23] A. Vilenkin, \PRD {\bf 30}, 569 (1984); \PRD {\bf 37}, 888
(1988); V.A. Rubakov, \PLB {\bf 148}, 280 (1984).

\endref
\bye